\documentclass[11pt]{article}
\pdfoutput=1
\usepackage{fullpage}
\usepackage{natbib}
\usepackage{amsmath,amsfonts,amssymb,amsthm}
\usepackage[blocks]{authblk}
\usepackage{algorithm}
\usepackage{algorithmic}
\usepackage{setspace}
\usepackage{enumitem}
\usepackage{graphicx}

\DeclareMathOperator{\diag}{diag}

\newcommand{\argmin}{\operatornamewithlimits{argmin}}

\renewcommand{\diag}{\operatornamewithlimits{diag}}

\newcommand{\Tr}{\mathbf{Tr}}

\makeatletter
\newif\if@borderstar
\def\bordermatrix{\@ifnextchar*{%
\@borderstartrue\@bordermatrix@i}{\@borderstarfalse\@bordermatrix@i*}%
}
\def\@bordermatrix@i*{\@ifnextchar[{\@bordermatrix@ii}{\@bordermatrix@ii[()]}}
\def\@bordermatrix@ii[#1]#2{%
\begingroup
\m@th\@tempdima8.75\p@\setbox\z@\vbox{%
\def\cr{\crcr\noalign{\kern 2\p@\global\let\cr\endline }}%
\ialign {$##$\hfil\kern 2\p@\kern\@tempdima & \thinspace %
\hfil $##$\hfil && \quad\hfil $##$\hfil\crcr\omit\strut %
\hfil\crcr\noalign{\kern -\baselineskip}#2\crcr\omit %
\strut\cr}}%
\setbox\tw@\vbox{\unvcopy\z@\global\setbox\@ne\lastbox}%
\setbox\tw@\hbox{\unhbox\@ne\unskip\global\setbox\@ne\lastbox}%
\setbox\tw@\hbox{%
$\kern\wd\@ne\kern -\@tempdima\left\@firstoftwo#1%
\if@borderstar\kern2pt\else\kern -\wd\@ne\fi%
\global\setbox\@ne\vbox{\box\@ne\if@borderstar\else\kern 2\p@\fi}%
\vcenter{\if@borderstar\else\kern -\ht\@ne\fi%
\unvbox\z@\kern-\if@borderstar2\fi\baselineskip}%
\if@borderstar\kern-2\@tempdima\kern2\p@\else\,\fi\right\@secondoftwo#1 $%
}\null \;\vbox{\kern\ht\@ne\box\tw@}%
\endgroup
}
\makeatother

\newcommand{\appendixone}{}
\newcommand{\onelineup}{}

\newcommand{\Pf}{Proof~of~}

\newtheorem{theorem}{Theorem}

\newtheorem{lemma}{Lemma}
\newtheorem{remark}{Remark}
\newtheorem{example}{Example}

\date{}

\title{Towards a sparse, scalable, and stably positive definite\\ (inverse) covariance estimator}

\author{Sang-Yun Oh}
\affil{Department of Statistics and Applied Probability, University of California \authorcr
Santa Barbara, California 93106-3110, U.S.A. \authorcr \texttt{syoh@pstat.ucsb.edu}}	

\author{Bala Rajaratnam}
\affil{Department of Statistics, Stanford University \authorcr
390 Serra Mall, Stanford, California 94305-4065, U.S.A. \authorcr \texttt{brajarat@stanford.edu}.}

\author{Joong-Ho Won}
\affil{Department of Statistics, Seoul National University \authorcr
1 Gwanak-ro, Gwanak-gu, Seoul 08826, Korea \authorcr \texttt{wonj@stats.snu.ac.kr}.}

\begin{document}
\maketitle

\begin{abstract}
\noindent
High dimensional covariance estimation and graphical models is a contemporary topic in statistics and machine learning having widespread applications. 
The problem is notoriously difficult in high dimensions as the traditional estimate is not even positive definite. 
An important line of research in this regard is to shrink the extreme spectrum of
 the covariance matrix estimators. 
A separate line of research in the literature has considered sparse inverse covariance estimation which in turn gives rise to graphical models. 
In practice, however, a sparse covariance or inverse covariance matrix which is simultaneously well-conditioned and at the same time computationally tractable is desired.  There has been little research at the confluence of these three topics. 
In this paper we consider imposing a condition number constraint to various types of losses used in covariance and inverse covariance matrix estimation.
This extends the approach by Won, Lim, Kim, and Rajaratnam (2013) on multivariate Gaussian log likelihood.
When the loss function can be decomposed as a sum of an orthogonally invariant function of the estimate and its inner product with a function of the sample covariance matrix, we show that a solution path algorithm can be derived, involving a series of ordinary differential equations. 
The path algorithm is attractive because it provides the entire family of estimates for all possible values of the condition number bound, at the same computational cost of a single estimate with a fixed upper bound. 
An important finding is that the proximal operator for the condition number constraint, which turns out to be very useful in regularizing loss functions that are not orthogonally invariant and may yield non-positive-definite estimates, can be efficiently computed by this path algorithm. 
As a concrete illustration of its practical importance, we develop an operator-splitting algorithm that imposes a guarantee of well-conditioning as well as positive definiteness to recently proposed convex pseudo-likelihood based graphical model selection methods 
(Zhang and Zou, 2014; Khare, Oh, and Rajaratnam, 2015).

\end{abstract}

\doublespacing
\section{Introduction}
We consider the problem of estimating the covariance matrix or its inverse (precision matrix) from $n$ independent copies of $p$-variate random vectors from some distribution. This estimation problem is becoming increasingly important in many statistical methods, from least squares reqression to graphical model selection. Applications include medical image analysis, genomics, and financial engineering, to name a few. In some applications (e.g., portfolio optimization, Gauss mixture clustering) overall risk properties of the covariance estimator are important; in others (e.g., graphical model selection), the sparsity pattern of the inverse covariance matrix is of critical interest. In any situation, the estimator should be symmetric, positive definite to be a valid (inverse) covariance matrix. It is also desirable that the ratios between the eigenvalues of the estimator are not too extremal, in order to reflect that the population covariance matrix describes a proper, non-degenerate $p$-dimensional distribution. In this paper, we call matrices that satisfy both conditions to be \emph{stably} positive definite.

Unfortunately, however, many estimators of covariance or inverse covariance matrix are not positive definite, let alone stably positive definite. It is well known that the sample covariance matrix
\begin{align}\label{eqn:samplecovariance}
S = \frac{1}{n}\sum_{i=1}^n (X_i-\bar{X})(X_i-\bar{X})^T,
\end{align}
where $X_i$ is the $i$th copy of the random vector, is merely positive \emph{semidefinite} when $n < p$. 
Some high-dimensional covariance matrix estimators based on structural sparsity assumptions may fail to be positive definite \citep*{fan2013large}; high-dimensional sparse inverse covariance matrix estimators based on maximum pseudo-likelihood principle \citep*{meinshausen2006high,peng2009partial,zhao2009composite,Khare2014, zhang2014sparse} 
may have negative eigenvalues, sometimes not even symmetric.

The main subject of study in this paper is the set of positive definite matrices with bounded condition numbers. The condition number of a positive definite matrix quantifies its degree of invertiblity, and is defined as the ratio of the largest to smallest eigenvalues of the matrix. Thus the set of interest can be formally written, for an upper bound $\kappa$,
\begin{align*}
\mathcal{C}_{\kappa} &= \{ \Omega: \Omega \succ 0, ~  \lambda_{\max}(\Omega)/\lambda_{\min}(\Omega) \le \kappa \} \\
	&= \{ \Omega: \exists u > 0, uI \preceq \Omega \preceq \kappa u I\},
\end{align*}
where $A \succ 0$ (\textit{resp}. $A \succeq 0$) denotes that matrix $A$ is positive definite (\textit{resp}. positive semidefinite), $A \preceq B$ means that $B-A \succeq 0$, and $\lambda_{\max}(A)$ and $\lambda_{\min}(A)$ refers to the maximum and minimum eigenvalues of $A$; the identity matrix is denoted by $I$. One should note that the set $\mathcal{C}_{\kappa}$ properly encodes the notion of stable positive definiteness.
If the (inverse) covariance matrix can be estimated constrained on $\mathcal{C}_{\kappa}$, then the estimator possesses the desired properties mentioned in the previous paragraph. Because $\Omega \in \mathcal{C}_{\kappa}$ implies that $\Omega^{-1}$ exists and $\Omega^{-1} \in \mathcal{C}_{\kappa}$, we do not distinguish estimation of the covariance matrix and estimation of the inverse covariance matrix too much; for the reason that will become apparent in the sequel, we use $\Omega$ to denote the inverse covariance matrix.
\citet*{Won2013} studied the set $\mathcal{C}_{\kappa}$ as a means to regularize high-dimensional Gaussian maximum likelihood covariance estimators. Their motivation is to impose numerical stability for inversion of the estimates, for instance to use with Markowitz-type portfolio optimization problems.
In this paper, we see this idea can be extended to a much general class of loss functions.

Now consider the estimation problem of the form
\begin{align}\label{eqn:orthogonalcondreg}
\begin{array}{ll}
	\text{minimize} & L(\Omega) - \Tr(\Omega f(S)) \\
	\text{subject~to} & \Omega \in \mathcal{C}_{\kappa},
\end{array}
\end{align}
where $L(\Omega)$ is convex; 
$S$ is the sample covariance matrix \eqref{eqn:samplecovariance}; and $f$ is a function that maps a symmetric matrix to a symmetric matrix of the same dimension.
Problem \eqref{eqn:orthogonalcondreg} includes many interesting cases:
\begin{enumerate}
	\item\label{itm:gaussian} Gaussian log likelihood: 
			$L(\Omega) = -\log\det\Omega$, $f(S) = -S$.
	\item\label{itm:nucnorm} Gaussian log likelihood with a-pair-of-nuclear-norms regularization \citep{Chi2013}:
			$L(\Omega) = -\log\det\Omega + \eta(\alpha\|\Omega\|_{*} + (1-\alpha)\|\Omega^{-1}\|_{*})$,
			$f(S) = -S$.
	\item\label{itm:quadratic} Quadratic loss:
			$L(\Omega) = (1/2)\|\Omega\|_F^2$, $f(S)=S$.
	\item\label{itm:concord} CONCORD loss \citep{Khare2014}:
			$L(\Omega) = -\log\det \Omega_D + (1/2)\Tr(\Omega S \Omega)$, $f(S)=0$, where $\Omega_D = \diag(\Omega_{11},\dotsc,\Omega_{pp})$.
	\item\label{itm:dtrace} D-trace loss \citep{zhang2014sparse}:
			$L(\Omega) = (1/2)\Tr(\Omega S \Omega)$, $f(S) = I$.
\end{enumerate}

Hence a characterization of the solution to \eqref{eqn:orthogonalcondreg} is of an utter interest. 
Cases \ref{itm:gaussian} -- \ref{itm:quadratic} are distinguished from the rest because in these cases $L(\Omega)$ is
orthogonally invariant, i.e., $L(Q^T\Omega Q) = L(\Omega)$ for any $Q$ such that $Q^TQ = I$, with an additional condition that $L(D) = \sum_{i=1}^p l_i(d_i)$, $l_i$ being closed convex, if $D=\diag(d_1,\ldots,d_p)$. 
For instance, 
$$
l_i(\lambda) = \begin{cases}
	-\log\lambda,  & \text{case~\ref{itm:gaussian}}, \\
	-\log\lambda + \eta(\alpha\lambda + (1-\alpha)\lambda^{-1}), & \text{case~\ref{itm:nucnorm}}, \\
	(1/2)\lambda^2, & \text{case~\ref{itm:quadratic}}.
\end{cases}
$$
In such cases, we can provide a complete characterization of the solution path of \eqref{eqn:orthogonalcondreg} as the parameter $\kappa$ varies from unity to infinity. 
Furthermore, we show that for many interesting cases, the entire solution path can be computed at the same cost (namely, in $O(p)$ operations) as that of finding the solution for a fixed $\kappa$. Thus the characterization of the solution path provides a huge computational advantage in solving \eqref{eqn:orthogonalcondreg} efficiently.

Cases \ref{itm:concord} and \ref{itm:dtrace} are pseudo-likelihood losses that arise in high-dimensional graphical model selection.
Orthogonal variance of $L(\Omega)$ in these cases prevents a direct application of the method mentioned in the previous paragraph. 
Nevertheless we can show that problem \eqref{eqn:orthogonalcondreg} with these losses can be efficiently solved by a scalable, Dykstra's alternating projection-type operator splitting method \citep{lange2013}, resulting in a sparse, stably positive definite covariance selection. This is because the orthogonal projection of a symmetric matrix to set $\mathcal{C}_{\kappa}$ has an almost closed form representation, a result that follows from Section \ref{sec:characterization}. In this sense, case \ref{itm:quadratic} bridges cases \ref{itm:gaussian} and \ref{itm:nucnorm} with cases \ref{itm:concord} and \ref{itm:dtrace}.

The rest of the paper is organized as follows. 
In Section \ref{sec:characterization}, we characterize the solution path for the orthogonally invariant cases as soultions of ordinary differential equations with respect to $\kappa$, 
introduce an efficient method to solve \eqref{eqn:orthogonalcondreg} for all values of $\kappa$ based on this observation. Explicit solutions to some cases introduced in this section are also provided.
In Section \ref{sec:splitting} we develop an alternating projection algorithm that solves the orthogonally variant cases scalably, and demonstrate that the algorithm provides stably positive semidefinite solutions to graphical model selection problems, without loosing the desired sparsity.
Section \ref{sec:conclusion} concludes this paper.
Some proofs of the results in the paper are given in the Appendix.

\section[Regularization Path]{Solution path for orthogonally invariant $L(\Omega)$}\label{sec:characterization}

We begin with the characterization of the solution to \eqref{eqn:orthogonalcondreg} for a fixed $\kappa$.
\begin{theorem}\label{thm:solution}
Suppose the spectral decomposition of $f(S)$ is given by $VDV^T$, $V^TV=VV^T=I$, $D=\diag(d_1,\dotsc,d_p)$, $d_1 \ge \dotsb \ge d_p$. Then, $\Omega^{\star} = V\Lambda^{\star}V^T$ minimizes \eqref{eqn:orthogonalcondreg}, where $\Lambda^{\star}=\diag(\lambda_1^{\star},\dotsc,\lambda_p^{\star})$ with
\begin{align}\label{eqn:optlambda}
\lambda_i^{\star} = \max( u^{\star}, \min(\tilde\lambda_i, \kappa u^{\star} ) ).
\end{align}
The $\tilde\lambda_i$ is the minimizer of $l_i(\lambda) - d_i\lambda$ in $\lambda \ge 0$.
Let $u_{\alpha,\beta}=\argmin_{u} l_{\alpha,\beta}(u)$ where
$$
l_{\alpha,\beta}(u) = \sum_{i=1}^{\alpha}l_{p-i+1}(u) - u\sum_{i=1}^{\alpha} d_{p-i+1}
	+ \sum_{i=\beta}^p l_{p-i+1}(\kappa u) - \kappa u \sum_{i=\beta}^p d_{p-i+1}
$$
for $\alpha \in \{1,\ldots,p-1\}$ and $\beta \in \{2,\ldots,p\}$. Then $u^{\star}$ can be chosen to equal to $u_{\alpha,\beta}$ for $(\alpha,\beta)$ satisfying the relation
$$
(u_{\alpha,\beta},v_{\alpha,\beta})
\in 
R_{\alpha,\beta} = \{ (u,v): \tilde\lambda_{p-\alpha+1} < u \le \tilde\lambda_{p-\alpha}, \tilde\lambda_{p-\beta+2} \le v < \tilde\lambda_{p-\beta+1} \},
\quad
v_{\alpha,\beta} = \kappa u_{\alpha,\beta}.
$$
Finding the pair $(\alpha,\beta)$ takes $O(p)$ time.
\end{theorem}
The proof is given in Appendix 1.
\begin{remark}
	This theorem subsumes \citet[Theorem 1]{Won2013} that corresponds to case \ref{itm:gaussian},
	and allows $f(S)$ to be indefinite or singular, i.e., $d_i \le 0$ for some $i$.
	Thus $\tilde\lambda_i=\infty$ or $\tilde\lambda_i=-\infty$ is allowed.
\end{remark}
\begin{remark}
	An insepection of the proof reveals that the problem reduces to determine 
	$$
	u^{\star} = \argmin_{u>0} \sum_{i=1}^p l_i(\lambda_i^{*}(u)) - d_i \lambda_i^{*}(u),
    $$
	where $\lambda_i^{*}(u) = \max(u,\min(\tilde\lambda_i,\kappa u) )$, i.e. a univariate minimization problem. Thus standard univariate optimization methods, e.g., bisection or golden search, can also be employed to find $u^{\star}$, subject to a tolerance level. The theorem says that it can be found \emph{exactly} within $O(p)$ operations.
\end{remark}

If $l_i$s are continuously differentiable, the $u_{\alpha,\beta}$ in Theorem \ref{thm:solution} can be found by solving the equation
\begin{align}\label{eqn:implicit}
\sum_{i=1}^{\alpha} l_{p-i+1}'(u) + \kappa\sum_{i=\beta}^p l_{p-i+1}'(\kappa u) = \sum_{i=1}^{\alpha}d_{p-i+1} + \kappa \sum_{i=\beta}^{p}d_{p-i+1}.
\end{align}
Then the implicit function theorem states that $u_{\alpha,\beta}=u_{\alpha,\beta}(\kappa)$ is a continous function of $\kappa$. Thus if the optimal $u^{\star}$ in \eqref{eqn:optlambda} satisifies $u^{\star}(\kappa) = u_{\alpha,\beta}(\kappa)$ so that $u_{\alpha,\beta}(\kappa),v_{\alpha,\beta} \in \mathbf{int} R_{\alpha,\beta}$ for some $\alpha,\beta$, 
where $\mathbf{int}A$ denotes the interior of a set $A$,
then a small change in $\kappa$ will not change $\alpha$ or $\beta$, i.e., $u^{\star}(\kappa+\Delta\kappa) = u_{\alpha,\beta}(\kappa+\Delta\kappa)$ and $(u_{\alpha,\beta}(\kappa+\Delta\kappa),v_{\alpha,\beta}(\kappa+\Delta\kappa)) \in \mathbf{int} R_{\alpha,\beta}$ for sufficiently small $\Delta\kappa$.
Thus the local solution path within $R_{\alpha,\beta}$ can be traced by solving \eqref{eqn:implicit} for continuously varying $\kappa$ subject to the condition $u^{\star}(\kappa) \in \mathbf{int} R_{\alpha,\beta}$.
If we further assume that $l_i$s are twice differentiable, this local path can be completely characterized by an ordinary differential equation: it is straighforward to derive
\begin{align}\label{eqn:ODE}
	\frac{du_{\alpha,\beta}}{d\kappa} &=
	\frac{\sum_{i=\beta}^p d_{p-i+1} - \sum_{i=\beta}^p l_{p-i+1}'(\kappa u) - \kappa u \sum_{i=\beta}^p l_{p-i+1}''(\kappa u)} 
	{\sum_{i=1}^{\alpha}l_{p-i+1}''(u) + \kappa^2 \sum_{i=\beta}^p l_{p-i+1}''(\kappa u)},
\end{align}
from which the curve $(u^{\star}(\kappa),v^{\star}(\kappa))$ within $R_{\alpha,\beta}$ can be determined.
\begin{example}
For case \ref{itm:gaussian}, we have
\begin{align}\label{eqn:gaussianODE}
\frac{d u_{\alpha,\beta}}{d\kappa} =  -\frac{(\alpha+p-\beta+1)\sum_{i=\beta}^{p} s_i}{(\sum_{i=1}^{\alpha} s_i + \kappa \sum_{i=\beta}^{p} s_i)^2},
\end{align}
where $s_i$ is the $i$th largest eigenvalue of $S$.
In this case \eqref{eqn:implicit} has an explicit solution
$$
u_{\alpha,\beta}(\kappa) = \frac{\alpha+p-\beta+1}{\sum_{i=1}^{\alpha} s_i + \kappa \sum_{i=\beta}^{p} s_i},
$$
which satisfies \eqref{eqn:gaussianODE}.
Furthermore, because
$$
\frac{d v_{\alpha,\beta}}{d\kappa} 
= \frac{(\alpha+p-\beta+1)\sum_{i=1}^{\alpha} l_i}{(\sum_{i=1}^{\alpha} s_i + \kappa \sum_{i=\beta}^{p} s_i)^2},
$$
it follows that
$$
\frac{d v_{\alpha,\beta}}{d u_{\alpha,\beta}}(\kappa) = -\frac{\sum_{i=1}^{\alpha} s_i}{\sum_{i=\beta}^{p} s_i},
$$
which is constant within $R_{\alpha,\beta}$. 
In other words, the solution path is piecewise linear in the $u$-$v$ plane. 
\end{example}
\begin{example}
For case \ref{itm:quadratic}, we have
$$
\frac{du_{\alpha,\beta}}{d\kappa} =
	\frac{\sum_{i=\beta}^p s_{p-i+1} - 2(p-\beta+1)\kappa u}
	{\alpha + \kappa^2(p-\beta+1)},
$$
whose general solution is given by
\begin{align}\label{eqn:quadraticpath}
u_{\alpha,\beta}(\kappa) = K \exp\left( \frac{\sum_{i=\beta}^p s_{p-i+1}}{\sqrt{\alpha(p-\beta+1)}}
		\tan^{-1}(\kappa\sqrt{\alpha^{-1}(p-\beta+1)}) + \log(\alpha + (p-\beta+1)\kappa^2) \right),
\end{align}
for some constant $K>0$.
\end{example}

Will the piecewise smooth solution path above be continuous as well? 
The concern is that at the boundary of the rectangle $R_{\alpha,\beta}$ where a small change of $\kappa$ indeed alters  $\alpha$ and/or $\beta$, there may be a jump in the path. The following lemma shows that this will not happen. 
\begin{lemma}\label{lemma:cont}
Suppose for some $\tilde{\kappa}$ with  $(u_{\alpha,\beta}(\tilde{\kappa}),v_{\alpha,\beta}(\tilde{\kappa})) \in \mathbf{int}R_{\alpha,\beta}$. Let $\bar{\kappa}=\sup\{\kappa: (u_{\alpha,\beta}(\kappa),v_{\alpha,\beta}(\kappa)) \in R_{\alpha,\beta} \}$. Then the point $(u_{\alpha,\beta}(\bar{\kappa}), v_{\alpha,\beta}(\bar{\kappa}) )$ 
coincides with either
$(u_{\alpha-1,\beta}(\bar{\kappa}), v_{\alpha-1,\beta}(\bar{\kappa}) ) \in R_{\alpha-1,\beta}$,
$(u_{\alpha,\beta+1}(\bar{\kappa}), v_{\alpha,\beta+1}(\bar{\kappa}) ) \in R_{\alpha,\beta+1}$,
or 
$(u_{\alpha-1,\beta+1}(\bar{\kappa}), v_{\alpha-1,\beta+1}(\bar{\kappa}) ) \in R_{\alpha-1,\beta+1}$ 
exclusively. 
\end{lemma}
The proof is given in  Appendix 1.

We have so far seen that the solution path is continuous and piecewise smooth, and how the curve pieces can be computed and traced. 
The remaining task is to determine the initial point the path. 
The initial point can be obviously chosen to the point that corresponds to $\kappa=1$, i.e., 
we need to find $\alpha$ and $\beta$ such that $(u_{\alpha,\beta}(1),v_{\alpha,\beta}(1)) \in R_{\alpha,\beta}$.
Note in this case that the closure of the desired $R_{\alpha,\beta}$ should intersect with the line $v=u$.
By construction, this occurs if and only if $\alpha=\beta-1$. 
Then, from \eqref{eqn:implicit} with $\kappa=1$, it follows that
\begin{align}\label{eqn:initialpoint}
	\sum_{i=1}^p l_i'(u) = \sum_{i=1}^p d_i = p \bar{d}, \quad \text{where} \quad \bar{d} = \frac{1}{p}\sum_{i=1}^p d_i,
\end{align}
and $u^{\star}(1)$ is found by solving this equation. 
In particular, if $l_i=l$ for $i=1,\dotsc,p$, then
$$
u^{\star}(1) = (l')^{-1}(\bar{d}),
$$
where $(l')^{-1}$ is the generalized inverse of $l'$, which exists because $l'$ is nondecreasing.
Thus for case \ref{itm:gaussian} we obtain $u^{\star}(1)=1/\bar{s}$, and for case \ref{itm:quadratic} we have $u^{\star}(1)=\bar{s}$.

Combining Lemma \ref{lemma:cont} and the above discussion, we are ready to fully describe the entire solution path, as stated in the following theorem.
\begin{theorem}\label{thm:path}
If $l_i$, $i=1,\dotsc,p$, are closed convex and twice differentiable,
the lower truncation value $u^\star(\kappa)$ for the optimal eigenvalue \eqref{eqn:optlambda} for problem \eqref{eqn:orthogonalcondreg}, together with the upper truncation value $v(\kappa)=\kappa u(\kappa)$ traces a piecewise smooth path on the $u$-$v$ plane as the regularization parameter $\kappa$ varies. 
The resulting solution path is given by the solutions of the series of ordinary differential equations \eqref{eqn:ODE}, and its slope is discontinuous only when it intersects the vertical lines $u = \tilde\lambda_1,\ldots,\tilde\lambda_p$ or horizontal lines $v=\tilde\lambda_1,\ldots,\tilde\lambda_p$. 
The initial point of this path is found by solving \eqref{eqn:initialpoint}, corresponding to $\kappa=1$.
This initial point as well as the entire path can be found in $O(p)$ operations (Algorithm \ref{alg:forward}).
\end{theorem}
\begin{algorithm}[h!]
\caption{Solution path algorithm for orthogonal $L$}
\label{alg:forward}
\onelineup 
\begin{tabbing}
{\small~1:}\enspace Set \= $\kappa_{\mathrm{new}} \gets 1$\\
{\small~2:}\enspace Find $u^\star_{\mathrm{new}} = v^\star_{\mathrm{new}}$ by solving \eqref{eqn:initialpoint}\\
{\small~3:}\enspace Find $\alpha$ such that $\tilde\lambda_{p-\alpha+1} < u^{\star}_{\mathrm{new}} \le \tilde\lambda_{p-\alpha}$; set $\beta \gets \alpha + 1$ \\
{\small~4:}\enspace Set $\mathcal{K} \gets \{\kappa_{\mathrm{new}} \}$, $\mathcal{I} \gets \{(\alpha,\beta)\}$ \\
{\small~5:}\enspace While ($\alpha \ge 1$ and $\beta \le p$) \\
{\small~6:}\enspace\> Compute $u_{\alpha,\beta}(\kappa)$ by solving \eqref{eqn:ODE} \\
{\small~7:}\enspace\> Set $\kappa_u \gets \inf\{\kappa \ge \kappa_{\mathrm{new}}: u_{\alpha,\beta}(\kappa)=\tilde\lambda_{p-\alpha+1}\}$ \\
{\small~8:}\enspace\> Set $\kappa_v \gets \inf\{\kappa \ge \kappa_{\mathrm{new}}: \kappa u_{\alpha,\beta}(\kappa)=\tilde\lambda_{p-\beta+1}\}$ \\
{\small~9:}\enspace\> Set $\kappa_{\mathrm{new}} \gets \min(\kappa_u,\kappa_v)$ \\
{\small 10:}\enspace\> $\mathcal{K} \gets \mathcal{K} \cup \{\kappa_{\mathrm{new}} \}$, $\mathcal{I} \gets \mathcal{I} \cup \{(\alpha,\beta)\}$ \\
{\small 11:}\enspace\> If $u_{\alpha,\beta}(\kappa_{\mathrm{new}}) = \tilde\lambda_{p-\alpha+1}$ then $\alpha \gets \alpha - 1$ \\
{\small 12:}\enspace\> If $\kappa_{\mathrm{new}} u_{\alpha,\beta}(\kappa_{\mathrm{new}}) = \tilde\lambda_{p-\beta+1}$ then $\beta \gets \beta + 1$ \\
{\small 13:}\enspace Return $\mathcal{K}, \mathcal{I}$
\end{tabbing}
\end{algorithm}
\onelineup 
\onelineup
\begin{proof}
Line 4 of Algorithm \ref{alg:forward} takes $O(p)$ operations. In the loop, either of the conditions in Lines 11 and 12 must be met for each iteration. Thus for each value of $\alpha=1,2,\ldots,p$, at most one value of $\beta \in \{1,2,\ldots,p\}$ is considered. This takes $O(p)$ time. 
\end{proof}
\begin{remark}
Algorithm \ref{alg:forward} terminates if $v^{\star}=\tilde\lambda_{p-r+1}$, where
$$
\tilde\lambda_1 \ge \dotsb \ge \tilde\lambda_r > \tilde\lambda_{r+1} = \dotsb = \tilde\lambda_p,
$$
understanding $\tilde\lambda_{p+1} = -\infty$.
This includes the case when $S$ is singular, i.e., 
$$
s_1 \ge \dotsb \ge s_r > 0 = s_{r+1} = \dotsb = s_p.
$$
\end{remark}

For case \ref{itm:gaussian}, using the fact that the solution path is piecewise linear in the $u$-$v$ plane, a simple geometric algorithm can be devised. This is shown in Algorithm \ref{alg:gaussian}.
\begin{algorithm}[h!]
\caption{Solution path algorithm for case \ref{itm:gaussian}}
\label{alg:gaussian}
\onelineup 
\begin{tabbing}
{\small~1:}\enspace Set \= $\kappa_{\mathrm{new}} \gets 1$, $u^\star_{\mathrm{new}} = v^\star_{\mathrm{new}} = 1/\bar s$\\
{\small~2:}\enspace Find $\alpha$ such that $l_{\alpha} > \bar l \ge l_{\alpha+1}$; set $\beta \gets \alpha + 1$ \\
{\small~3:}\enspace Set $\mathcal{K} \gets \{\kappa_{\mathrm{new}}\}$, $\mathcal{U} \gets \{u^\star\}$, $\mathcal{V}= \{v^\star\}$ \\
{\small~4:}\enspace While ($\alpha \ge 1$ and $\beta \le p$) \\
{\small~5:}\enspace\> $t \leftarrow -(\sum_{i=1}^{\alpha}l_i)/(\sum_{i=\beta}^p l_i)$ \\
{\small~6:}\enspace\> $\bar{R}_{\alpha,\beta} \leftarrow \left\{(u,v) : 1/l_{\alpha} \le u \le 1/l_{\alpha+1}~\mbox{and}~1/l_{\beta-1} \le v \le 1/l_{\beta} \right\}$ \\
{\small~7:}\enspace\> $(u^*,v^*) \leftarrow$ intersection \= between line passing $\left(u^\star_{\mathrm{new}}, v^\star_{\mathrm{new}}\right)$ of slope $t$ \\
	\enspace\>\> and boundary of $\bar{R}_{\alpha,\beta}$, with $u^* < u^\star_{\mathrm{new}}$ \\
{\small~8:}\enspace\> $\kappa_{\mathrm{new}} \leftarrow v^*/u^*$, $u^\star_{\mathrm{new}} \leftarrow u^*$, $v^\star_{\mathrm{new}} \leftarrow v^*$ \\
{\small~9:}\enspace\> $\mathcal{K} \gets \mathcal{K} \cup \{\kappa_{\mathrm{new}}\}$, $\mathcal{U} \gets \mathcal{U} \cup \{u^\star_{\mathrm{new}}\}$, $\mathcal{V} \gets \mathcal{V} \cup \{ v^\star_{\mathrm{new}} \}$ \\
{\small 10:}\enspace\> If $u^* = 1/s_{\alpha}$ then $\alpha \gets \alpha - 1$ \\
{\small 11:}\enspace\> If $v^* = 1/s_{\beta}$ then  $\beta \gets \beta + 1$ \\
{\small 12:}\enspace Return $\mathcal{K}, \mathcal{U}, \mathcal{V}$
\end{tabbing}
\end{algorithm}
\onelineup 
\onelineup
\onelineup

\section{Solution procedure for orthogonally variant $L(\Omega)$}\label{sec:splitting}

With an additional sparsity-incuding penalty, problem \eqref{eqn:orthogonalcondreg} can be compactly written
\begin{align}\label{eqn:concordcondreg}
\begin{array}{ll}
\text{minimize} &
h_1(\Omega) + h_2(\Omega) \\
\text{subject~to} & 
\Omega \in \mathcal{C}_{\kappa},
\end{array}
\end{align}
where $h_1(\Omega)=L(\Omega)-\Tr(\Omega f(S)$ and $h_2(\Omega) = \mu|\Omega|_1 = \mu \sum_{i<j} |\Omega_{ij}|$. 
To be specific,
$$
h_1(\Omega) =
\begin{cases}
	-\log\det \Omega_D + (1/2)\Tr(\Omega S \Omega), & \text{case~\ref{itm:concord}}, \\
	(1/2)\Tr(\Omega S \Omega) - \Tr(\Omega), & \text{case~\ref{itm:dtrace}}.
\end{cases}
$$
Problem \eqref{eqn:concordcondreg} can be equivalently written
\begin{align*}
\text{minimize} ~
h_1(\Omega) + h_2(\Omega) + \mathcal{I}_{\mathcal{C}_{\kappa}}(\Omega),
\end{align*}
where 
$$
\mathcal{I}_{\mathcal{C}_{\kappa}}(\Omega) =
\begin{cases} 
	0, & \Omega \in \mathcal{C}_{\kappa} \\
	+\infty, & \text{otherwise}.
\end{cases}
$$
is the indictor function of the set $\mathcal{C}_{\kappa}$.
Because both $h_1$ and $h_2$ are not orthogonally invariant, it is not obvious how to handle this spectral constraint set efficiently. 
The key idea here is to utilize the fact that the proximal operator of the indicator function $\mathcal{I}_{\mathcal{C}_{\kappa}}$, that is, the orthogonal projection to $\mathcal{C}_{\kappa}$, is efficiently computed using Algorithm \ref{alg:forward}.
For $X \in \mathbb{S}^{p}$, where $\mathbb{S}^p$ is the space of $p\times p$ symmetric matrices, 
the proximal operator is defined as follows.
\begin{align}\label{eqn:projection}
\mathcal{P}_{\mathcal{C}_{\kappa}}(X) = \argmin_{\tilde{X} \in \mathbb{S}^{p}} \mathcal{I}_{\mathcal{C}_{\kappa}}(\tilde{X}) + \frac{1}{2t}\| \tilde{X} - X \|_F^2, \quad t > 0.
\end{align}
The optimization problem involved in the right hand side of \eqref{eqn:projection} is
$$
\begin{array}{ll}
	\text{minimize} & (1/2)\|\tilde{X} - X\|_F^2 \\
	\text{subject~to} & \tilde{X} \in \mathcal{C}_{\kappa},
\end{array}
$$
i.e., case \ref{itm:quadratic}. Thus, Algorithm \ref{alg:forward} gives the entire solution to \eqref{eqn:projection} for all $\kappa \ge 1$ in $O(p)$ operations, with the smooth pieces has a closed form given in \eqref{eqn:quadraticpath}, given the spectral decomposition of $X$.

Now \eqref{eqn:concordcondreg} can be solved by using Dykstra's alternating projection algorithm \citep[Ch. 15]{lange2013}:
\begin{align}
	\Omega^{(k+1/2)} & := \argmin_{\Omega \in \mathbb{S}^p} h_1(\Omega) + h_2(\Omega) + (1/2)\| \Omega - \bar\Omega^{(k)} \|_F^2 \label{eqn:dykstra} \\
	\bar\Omega^{(k+1/2)} & := 2\Omega^{(k+1/2)} - \bar\Omega^{(k)} \nonumber \\
	\Omega^{(k+1)} & := \mathcal{P}_{\mathcal{C}_{\kappa}}(\bar\Omega^{(k+1/2)})\nonumber \\
	\bar\Omega^{(k+1)} & := \bar\Omega^{(k)} + \Omega^{(k+1)} - \Omega^{(k+1/2)}, \nonumber
\end{align}
which is an instance of the Douglas-Rachford operator splitting algorithm \citep{eckstein1992douglas};
converges is guaranteed if $h_1(\Omega) + h_2(\Omega)$ is closed convex, which holds for cases \ref{itm:concord} and \ref{itm:dtrace}.

For case \ref{itm:concord}, the subproblem \eqref{eqn:dykstra} is to solve
$$
\text{minimize} \quad 
 -\log\det \Omega_D + (1/2)\Tr(\Omega (S+(1/2)I) \Omega) - \Tr(\Omega \bar\Omega^{(k)}) + \mu | \Omega |_1,
$$
which is yet another CONCORD problem. This problem can be efficiently solved via the block coordinate descent \citep{Khare2014}, or proximal gradient methods \citep*{oh2014optimization}.

For case \ref{itm:dtrace}, \eqref{eqn:dykstra} reduces to a lasso program \citep{Tibshirani1996}:
$$
\text{minimize} \quad 
 (1/2)\Tr(\Omega (S+(1/2)I) \Omega) - \Tr(\Omega(I+\bar\Omega^{(k)})) + \mu | \Omega |_1,
$$
which can again be efficiently solved via proximal gradient methods \citep{beck2009fast}.

\begin{figure}[ht]
\centering
	\includegraphics[width=0.75\textwidth]{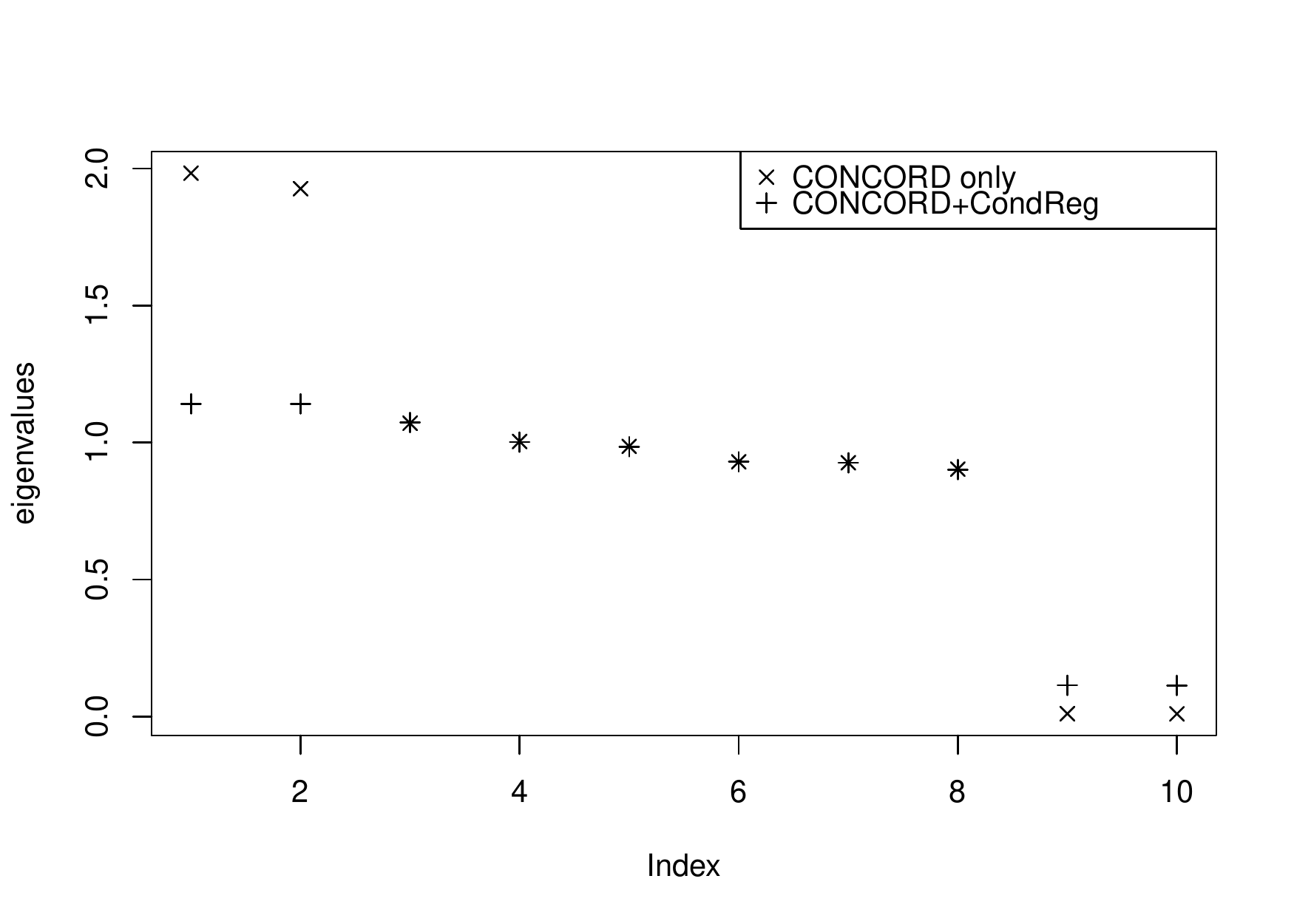}
\caption{Distribution of the eigenvalues of the CONCOND-only inverse covariance matrix estimate ($\times$), and CONCORD with condition number regularization ($+$).}\label{fig:eigdistribution}
\end{figure}

\paragraph{Illustration}
To illustrate the effect of the condition number regularization, we generated $n=200$ samples from $p=10$ dimensional multivariate normal distribution with zero mean and inverse covariance matrix $\Omega$ such that $\Omega_{ii}=1$ for $i=1,\dotsc,p$ and $\Omega_{15}=\Omega_{51}=\Omega_{26}=\Omega_{62}=.99$. We compared the estimated $\Omega$ obtained using the CONCORD-ISTA algorithm \citep{oh2014optimization} with sparisty level $\mu=0.1$ and that using the alternating projection algorithm of this section, where the upper bound for the condition number is set to 10 and the same CONCORD-ISTA is used for the subproblem \eqref{eqn:dykstra}. With the tolerance for the relative change of the estimates set as $1\times10^{-6}$ (the meanings of the relative change are not the same between these two, though), the former terminated within 1000 iterations, and the latter within 503 iterations, where the inner CONCORD-ISTA is ran up to 100 iterations for each outer iteration.
Both methods gave a similar sparsity pattern for the estimates (Figure \ref{fig:sparsity}). However, the inverse covariance matrix obtained using the CONCORD loss only is on the vicinity of singularity, with the minimum eigenvalue of 0.0102. The maximum eigenvalue was 1.98, giving the condition number of 194. On the other hand, the CONCORD loss combined with the condition number regularization yielded the minimum eigenvalue of 0.114, more than 10 times greater than the pseudo-likelihood-only counterpart, while the maximum eigenvalue was moderately reduced to 1.14. (Thus the condition number bound of 10.0 was retained.) The eigenvalue distributions of both cases are shown in Figure \ref{fig:eigdistribution}.

\begin{figure}[ht]
\centering
\footnotesize
$$
\begin{bmatrix}
  1.00 &      &      &      & 0.99 &      &      &      &      &      \\ 
       & 1.00 &      &      &      & 0.99 &      &      &      &      \\ 
       &      & 1.00 &      &      &      &      &      &      &      \\ 
       &      &      & 1.00 &      &      &      &      &      &      \\ 
  0.99 &      &      &      & 1.00 &      &      &      &      &      \\ 
       & 0.99 &      &      &      & 1.00 &      &      &      &      \\ 
       &      &      &      &      &      & 1.00 &      &      &      \\ 
       &      &      &      &      &      &      & 1.00 &      &      \\ 
       &      &      &      &      &      &      &      & 1.00 &      \\ 
       &      &      &      &      &      &      &      &      & 1.00 \\ 
\end{bmatrix}
$$
\\\normalsize (a) True $\Omega$\\
\footnotesize
$$
\begin{bmatrix}
  1.03 & 0.00 & 0.00 & -0.00 & 1.01 & -0.01 & 0.01 & -0.00 & 0.01 & 0.00 \\ 
  0.00 & 0.99 & 0.00 & -0.00 & 0.01 & 0.98 & 0.00 & 0.00 & 0.00 & -0.00 \\ 
  0.00 & 0.00 & 1.04 & -0.01 & -0.00 & -0.00 & -0.06 & 0.00 & -0.01 & 0.00 \\ 
 -0.00 & -0.00 & -0.01 & 0.93 & 0.00 & 0.00 & 0.00 & 0.00 & 0.00 & 0.00 \\ 
  1.01 & 0.01 & -0.00 & 0.00 & 1.02 & 0.00 & 0.00 & 0.00 & 0.00 & 0.01 \\ 
 -0.01 & 0.98 & -0.00 & 0.00 & 0.00 & 0.99 & 0.00 & -0.01 & -0.03 & 0.00 \\ 
  0.01 & 0.00 & -0.06 & 0.00 & 0.00 & 0.00 & 0.93 & 0.00 & 0.00 & 0.00 \\ 
 -0.00 & 0.00 & 0.00 & 0.00 & 0.00 & -0.01 & 0.00 & 0.93 & 0.02 & 0.00 \\ 
  0.01 & 0.00 & -0.01 & 0.00 & 0.00 & -0.03 & 0.00 & 0.02 & 1.00 & 0.00 \\ 
  0.00 & -0.00 & 0.00 & 0.00 & 0.01 & 0.00 & 0.00 & 0.00 & 0.00 & 0.99 \\ 
\end{bmatrix}
$$
\\\normalsize (b) CONCORD estimate\\
\footnotesize
$$
\begin{bmatrix}
  0.63 & 0.00 & 0.00 & -0.00 & 0.51 & -0.00 & 0.00 & -0.00 & 0.01 & 0.00 \\ 
  0.00 & 0.63 & 0.00 & -0.00 & 0.00 & 0.51 & 0.00 & 0.00 & 0.01 & -0.00 \\ 
  0.00 & 0.00 & 1.04 & -0.01 & -0.00 & -0.00 & -0.06 & 0.00 & -0.01 & 0.00 \\ 
 -0.00 & -0.00 & -0.01 & 0.93 & 0.00 & 0.00 & 0.00 & 0.00 & 0.00 & 0.00 \\ 
  0.51 & 0.00 & -0.00 & 0.00 & 0.63 & 0.00 & -0.00 & 0.00 & 0.00 & 0.01 \\ 
 -0.00 & 0.51 & -0.00 & 0.00 & 0.00 & 0.63 & 0.00 & -0.00 & -0.02 & 0.00 \\ 
  0.00 & 0.00 & -0.06 & 0.00 & -0.00 & 0.00 & 0.93 & 0.00 & 0.00 & 0.00 \\ 
 -0.00 & 0.00 & 0.00 & 0.00 & 0.00 & -0.00 & 0.00 & 0.93 & 0.02 & 0.00 \\ 
  0.01 & 0.01 & -0.01 & 0.00 & 0.00 & -0.02 & 0.00 & 0.02 & 1.00 & 0.00 \\ 
  0.00 & -0.00 & 0.00 & 0.00 & 0.01 & 0.00 & 0.00 & 0.00 & 0.00 & 0.99 \\ 
\end{bmatrix}
$$
\\\normalsize (c) CONCORD estimate with an upper bound on condition number\\
\caption{Illustration of the effect of the condition number regularization on the CONCORD pseudo-likelihood graphical model section.}\label{fig:sparsity}
\end{figure}

\section{Conclusion}\label{sec:conclusion}
We have considered imposing a condition number constraint to regularize the estimator of the covariance of inverse covariance matrix of a population distribution under various loss criteria. For the losses that consists of an orthogonally invariant term and an inner product with a function of the sample covariance matrix, the problem reduces essentially that of the eigenvalues of the estimator, and the entire solution path with respect to the degree of condition number regularization can be obtained. If the involved ordinary differential equation admits a closed form solution, then the path can be obtained at the same cost as finding the estimator for a fixed regularization parameter. For other losses, an operator splitting scheme can be employed to find the estimator, hence the problem is scalable. At the core of this scheme lies the fact that the projection operator to the set of matrices with bounded conditio numbers allows path solutions, due to its orthogonal invariance. 

The most expensive part in computing the solution paths is the spectral decomposition.
As noted by \citet{Chi2013},
randomized algorithms such as random projection to lower dimensional subspaces may provide a computational relief \citep{mahoney2011}. These approaches incurs a small loss in accuracy, thus a possible research direction is to handle inexact solutions to the optimization subprolems in the alternating projection algorithm properly.

\appendix

\appendixone
\section*{Appendix 1}\label{app:proofs}

\begin{proof}[\Pf Theorem~\ref{thm:solution}]
First note that both $L(\Omega)$ and $\mathcal{C}_{\kappa}$ are orthogonally invariant, hence depends only on the eigenvalues of $\Omega$. Suppose the spectral decomposion of $\Omega$ is $U\Lambda U^T$, $U^TU=UU^T=I$, $\Lambda=\diag(\lambda_1,\dotsc,\lambda_p)$, $\lambda_1 \ge \dotsb \ge \lambda_p$. 
For the trace part of the objective, the von Neumann-Fan inequality \citep[Appendix A.4]{mirsky1975trace,farrell1985,lange2013} 
asserts that
$$
\Tr(\Omega f(S)) \le \Tr(\Lambda D) = \sum_{i=1}^p \lambda_i d_i,
$$
with equality if and only if $V=U$. Thus problem \eqref{eqn:orthogonalcondreg} reduces to a $p+1$-variate problem
\begin{align}\label{eqn:eigenvalues}
\begin{array}{ll}
	\text{minimize} & \sum_{i=1}^p l_i(\lambda_i) - d_i \lambda_i \\
	\text{subject~to} & u \le \lambda_i \le \kappa u, \quad i=1,\dotsc,p, \\
	~				  & \lambda_1 \ge \dotsb \ge \lambda_p,
\end{array}
\end{align}
where the variables are $\lambda_1, \dotsc, \lambda_p$ and $u$.
The last order constraint can be removed, because of the following. 
Without the order constraint, for a fixed $u > 0$, the reduced problem \eqref{eqn:eigenvalues} becomes separable in $\lambda_i$; it suffices to solve
\begin{align}\label{eqn:separable}
\begin{array}{ll}
	\text{minimize} & l_i(\lambda_i) - d_i\lambda_i \\
	\text{subject~to} & u \le \lambda_i \le \kappa u 
\end{array}
\end{align}
for each $i = 1,\dotsc, p$. 
Convexity of the objective in \eqref{eqn:separable} ensures that the minimum is attained at
$$
\lambda_i^{*}(u) = \max(u, \min(\tilde\lambda_i, \kappa u) ),
$$
where $\tilde\lambda_i = \argmin_{\lambda} l_i(\lambda) - d_i \lambda$.
The optimality condition for $\tilde\lambda_i$ is given by
$$
d_i \in \partial l_i(\tilde\lambda_i)
\iff
\tilde\lambda_i \in \partial g^*(d_i),
$$
where $\partial f(x)$ denotes the subdifferential of $f$ at $x$, and $g^*(v) = \sup \langle \lambda,u \rangle - g(\lambda)$, the convex conjuate of $g(\lambda)$.
Monotoniciy of the subdifferential operator ensures that $\tilde\lambda_i$s perserve the order of $d_i$s, i.e., 
$\tilde\lambda_1 \ge \dotsb \ge \tilde\lambda_p$.
It follows that $\lambda_1^{*}(u) \ge \dotsb \ge \lambda_p^{*}(u)$, hence \eqref{eqn:eigenvalues} reduces to a univariate minimization problem over $u$
\begin{align}\label{eqn:univariate}
\text{minimize} ~ \sum_{i=1}^p l_i(\lambda_i^{*}(u)) - d_i\lambda_i^{*}(u).
\end{align}
The solution to \eqref{eqn:univariate}, $u^{\star}$, must satisify
$$
\lambda_{p-i+1}^{*}(u^{\star}) = 
	\begin{cases}
	u^{\star}, & i=1,\dotsc, \alpha^{\star}, \\
	\tilde\lambda_i, & i=\alpha^{\star}+1, \dotsc, \beta^{\star}-1, \\
	\kappa u^{\star}, &  i=\beta^{\star}, \dotsc, p,
	\end{cases}
$$
where $\alpha^{\star}$ and $\beta^{\star}$ are such that $\tilde\lambda_{p-\alpha^{\star}+1} < u \le \tilde\lambda_{p-\alpha^{\star}}$
and $\tilde\lambda_{p-\beta^{\star}+2} \le \kappa u^{\star} < \tilde\lambda_{p-\beta^{\star}+1}$.
To find $u^{\star}$, 
for $\alpha \in \{1,\ldots,p-1\}$ and $\beta \in \{2,\ldots,p\}$, define
$$
\lambda_{p-i+1}^{\alpha,\beta}(u) = 
	\begin{cases}
	u, & i=1,\dotsc, \alpha, \\
	\tilde\lambda_i, & i=\alpha+1, \dotsc, \beta-1, \\
	\kappa u, &  i=\beta, \dotsc, p,
	\end{cases}
$$
and
$$
u_{\alpha,\beta} = \argmin_u \sum_{i=1}^p l_{p-i+1}(\lambda_{p-i+1}^{\alpha,\beta}(u)) - d_{p-i+1}\lambda_{p-i+1}^{\alpha,\beta}(u)
	= \argmin_u l_{\alpha,\beta}(u).
$$
By construction, $u_{\alpha,\beta}$ coincides with $u^\star$ if and only if
\begin{align}\label{eqn:alphabeta}
	\tilde\lambda_{p-\alpha+1} < u_{\alpha,\beta} \le \tilde\lambda_{p-\alpha}
	\quad \text{and} \quad
	\tilde\lambda_{p-\beta+2} \le \kappa u_{\alpha,\beta} < \tilde\lambda_{p-\beta+1}.
\end{align}
or $(u_{\alpha,\beta}, \kappa u_{\alpha,\beta}) \in R_{\alpha,\beta}$.
Because $R_{\alpha,\beta}$s partition the $u$-$v$ plane into $(p+2)^2$ regions
and $(u_{\alpha,\beta}, \kappa u_{\alpha,\beta})$ is on the line $v = \kappa u$, an obvious algorithm to find the pair $(\alpha,\beta)$ that satisfies the condition \eqref{eqn:alphabeta} is to keep track of the rectangles $R_{\alpha,\beta}$ that intersect this line.
To see that this algorithm takes $O(p)$ operations, start from the origin of the $u$-$v$ plane, increase $u$ and $v$ along the line $v = \kappa u$. Since $\kappa \ge 1$, if the line intersects $R_{\alpha,\beta}$, then the next intersection occurs in one of the three rectangles: $R_{\alpha+1,\beta}$, $R_{\alpha,\beta+1}$, and $R_{\alpha+1,\beta+1}$. Therefore after finding the first intersection (which is on the line $u=\tilde\lambda_1$), the search requires at most $2p$ tests to satisfy condition \eqref{eqn:alphabeta}. Finding the first intersection takes at most $p$ tests.
\end{proof}

\begin{proof}[\Pf Lemma~\ref{lemma:cont}]
Increase $\kappa$ from $\bar{\kappa}$. 
Suppose the curve passing the point $(u_{\alpha^{\star},\beta^{\star}}(\tilde{\kappa}),v_{\alpha^{\star},\beta^{\star}}(\tilde{\kappa}))$ 
meets the left side (but not inclusive) $\{(u,v): u=\tilde\lambda_{p-\alpha+1}\}$ of 
$R_{\alpha,\beta}$ 
before it meets the upper side (also not inclusive) 
$\{(u,v): v=\tilde\lambda_{p-\beta+1}\}$. 
Then, taking the limit of both sides of \eqref{eqn:implicit} as $\kappa \nearrow \bar\kappa$, and by continuity of $u_{\alpha,\beta}(\kappa)$, we have
\begin{align}\label{eqn:leftboundary}
\sum_{i=1}^{\alpha} l_{p-i+1}'(\tilde\lambda_{p-\alpha+1}) + \bar\kappa\sum_{i=\beta}^p l_{p-i+1}'(\bar\kappa \tilde\lambda_{p-\alpha+1}) = \sum_{i=1}^{\alpha}d_{p-i+1} + \bar\kappa \sum_{i=\beta}^{p}d_{p-i+1}.
\end{align}
Optimality of $\tilde\lambda_{p-\alpha+1}$ (see \eqref{eqn:separable}) and continuity of $l_{p-i+1}'$ asserts that
$$
l_{p-i+1}'(\tilde\lambda_{p-\alpha+1}) = d_{p-\alpha+1}.
$$
Thus \eqref{eqn:leftboundary} is equivalent to
$$
\sum_{i=1}^{\alpha-1} l_{p-i+1}'(\tilde\lambda_{p-\alpha+1}) + \bar\kappa\sum_{i=\beta}^p l_{p-i+1}'(\bar\kappa \tilde\lambda_{p-\alpha+1}) = \sum_{i=1}^{\alpha-1}d_{p-i+1} + \bar\kappa \sum_{i=\beta}^{p}d_{p-i+1}.
$$
In other words,
$$
\tilde\lambda_{p-\alpha+1}
=u_{\alpha-1,\beta}(\bar{\kappa})
$$
and 
$(u_{\alpha,\beta}(\bar{\kappa}), v_{\alpha,\beta}(\bar{\kappa}) ) = (u_{\alpha-1,\beta}(\bar{\kappa}), v_{\alpha-1,\beta}(\bar{\kappa}) ) \in R_{\alpha-1,\beta}$. 
If the curve meets the upper side before the left side of $R_{\alpha,\beta}$, we have
\begin{align*}
\sum_{i=1}^{\alpha} l_{p-i+1}'(\tilde\lambda_{p-\beta+1}/\bar\kappa) + \bar\kappa\sum_{i=\beta}^p l_{p-i+1}'(\tilde\lambda_{p-\beta+1}) &= \sum_{i=1}^{\alpha}d_{p-i+1} + \bar\kappa \sum_{i=\beta}^{p}d_{p-i+1}, \\
l_i'(\tilde\lambda_{p-\beta+1}) &= d_{p-\beta+1},
\end{align*}
and thus
$$
\sum_{i=1}^{\alpha} l_{p-i+1}'(\tilde\lambda_{p-\beta+1}/\bar\kappa) + \bar\kappa\sum_{i=\beta+1}^p l_{p-i+1}'(\tilde\lambda_{p-\beta+1}) = \sum_{i=1}^{\alpha}d_{p-i+1} + \bar\kappa \sum_{i=\beta+1}^{p}d_{p-i+1}
$$
to have
$(u_{\alpha,\beta}(\bar{\kappa}), v_{\alpha,\beta}(\bar{\kappa}))
=(u_{\alpha,\beta+1}(\bar{\kappa}), v_{\alpha,\beta+1}(\bar{\kappa}) ) \in R_{\alpha,\beta+1}$.
The final case, that the curve meets the upper left corner of $R_{\alpha,\beta}$,
is the combination of previous two cases, and it follows that
$(u_{\alpha,\beta}(\bar{\kappa}), v_{\alpha,\beta}(\bar{\kappa}))
=(u_{\alpha-1,\beta+1}(\bar{\kappa}), v_{\alpha-1,\beta+1}(\bar{\kappa}) ) \in R_{\alpha-1,\beta+1}$.
\end{proof}

%
%
%
%
%

\bibliographystyle{chicago}

\bibliography{library}
\end{document}